\documentclass[useAMS,usenatbib]{mn2e}

\usepackage{graphicx,times}

\def\aj{{AJ}}
 
\def\mnras{{MNRAS}}
\def\msun{M_\odot}

\def\gtorder{\mathrel{\raise.3ex\hbox{$>$}\mkern-14mu
             \lower0.6ex\hbox{$\sim$}}}

\title[Monte Carlo Simulations of Star Clusters - VI. The globular
  cluster NGC 6397]{Monte Carlo Simulations of Star Clusters -
  VI.  The globular cluster NGC 6397}
\author[M. Giersz and D.C. Heggie]{Mirek Giersz$^{1}$\thanks{E-mail:
mig@camk.edu.pl (MG); d.c.heggie@ed.ac.uk (DCH)} and Douglas
 C. Heggie$^{2}$\thanks{On leave from the University
  of Edinburgh}
\\
$^{1}$Nicolaus Copernicus Astronomical Centre, Polish Academy of Sciences, ul. Bartycka 18, 00-716 Warsaw, Poland\\
$^2$Yukawa Institute for Theoretical Physics, Kyoto University,
Oiwake-cho, Kitashirakawa, Sakyo-ku, Kyoto, Japan}
\begin{document}

\date{Accepted \ldots. Received \ldots; in original form \ldots}

\pagerange{\pageref{firstpage}--\pageref{lastpage}} \pubyear{2002}

\maketitle

\label{firstpage}

\begin{abstract}
We describe Monte Carlo models for the dynamical evolution of the
nearby globular cluster NGC 6397.  The code includes treatments of two-body
relaxation, most kinds of three- and four-body interactions involving primordial
binaries and those formed dynamically, the Galactic tide, and the
internal evolution of both single and binary stars.  We arrive at a
set of initial parameters for the cluster which, after 12Gyr of
evolution, gives a model with a fairly satisfactory match to the surface
brightness profile, the velocity dispersion profile, and the
luminosity function in two fields.  We describe in particular those
aspects of the evolution which distinguish this cluster from M4, which
has a roughly similar mass and Galactocentric distance, but a
qualitatively different surface brightness profile.  Within the
limitations of our modelling, we conclude that
{ the most plausible explanation for the difference is fluctuations:
both clusters are post-collapse objects, but sometimes have resolvable
cores and sometimes not.}
\end{abstract}


\section{Introduction}

This is the third paper in which we aim to construct a dynamical
evolutionary model of a specific Galactic globular cluster, following
$\omega$ Cen \citep{GH2003} and M4 \citep{HG2008}.  We are motivated
by a variety of questions:  How did globular clusters come to be as
they are?  What was the initial mass function of their stars, and the
abundance of their binaries?  Where should we look for signs of the
dynamical evolution of their binary population?  

So far, our most interesting finding concerned M4, where we were
surprised to find that our model had exhibited core collapse about
4Gyr ago, even though this cluster is not a classic core collapse
cluster, to judge by its surface brightness profile \citep{Tr1995}.
Now we turn to an object which is almost the twin of M4, i.e. the
southern globular cluster NGC6397.  As we shall see in
Sec.\ref{sec:data} these two clusters have much the same mass,
Galactic orbit, binary fraction, etc.  The one feature where they
differ strikingly is the surface brightness profile:  NGC6397 {\sl
does} exhibit a collapsed core.  Because of their other similarities,
conventional explanations for this difference are untenable.  In this
paper our chief aim is to provide a plausible explanation of this
contrast between these two clusters.

Our technique is
fundamentally a Monte Carlo simulation method in a style originally
devised by \citet{He1971}, but developed afresh by
\citet{Gi1998,Gi2001,Gi2006}, partly following ideas developed by
\citet{St1982,St1986}.  It now incorporates all the main processes
affecting the long-term evolution of globular star clusters, including
the Galactic tide and the internal evolution of single and binary stars, not to
mention dynamical interactions between these populations.  It has been
tested against results of $N$-body simulations of the rich open
cluster M67 \citep{GHH2008}, and at present such a code is an unmatched tool for
the evolutionary modelling of  globular clusters.

This is by no means the first time dynamical models have been
constructed for this cluster.  Almost all previous modelling, however,
was concerned with the construction of an equilibrium model, such as a
King model.  Such models do not readily lend themselves to
consideration of dynamical evolution.  Exceptions are the $N$-body
models described in \citet{Ri2008,Hu2008,Da2008a} and the Fokker-Planck
modelling of \citet{Dr1993,Dr1995}.  Together with our Monte Carlo
method, these techniques offer three ways in which  the effects of
dynamical evolution on a globular cluster may be assessed. 

 The
$N$-body models are essentially free of simplifying dynamical
assumptions, except that a globular cluster like NGC6397 is too rich
{ for its entire evolution} to be modelled directly.  Therefore assumptions are needed when
$N$-body models are applied to such objects.  For example,
\citet{Ri2008} and \cite{Hu2008} observe that a particular $N$-body
model, at an age of 16Gyr, has a mass function which conforms
approximately to that observed in a certain field in NGC6397, though
the model has a mass at 16Gyr of only about $10^4\msun$, which is considerably
smaller than estimates for this cluster (Tables \ref{tab:data},
\ref{tab:ic-values}).  As these authors emphasise, their $N$-body model is
not a model of NGC6397, but is intended to allow estimates of the
effect of dynamical evolution on the spatial distribution of binaries,
and so on.

The Fokker-Planck models of \citet{Dr1995} are essentially scale-free,
but the paper includes an exhaustive discussion of how the scales and
initial parameters of the models are constrained by fitting to 
profiles of surface density and velocity dispersion, and to ground
based mass functions at three radii.  

Our own approach combines some of the advantages of $N$-body and
Fokker-Planck techniques.  Similar to Drukier's, our aim is to construct a
dynamical evolutionary model which provides a satisfactory fit to
surface brightness and velocity dispersion profiles, and to luminosity
functions.  But our method of simulation is a star-by-star method,
like the $N$-body technique, and it shares with \citet{Hu2008} the use
of synthetic modelling of the internal evolution of both single and
binary stars.  Our advantage over the $N$-body modelling is the use of
a realistic number of particles (which depends in turn on speed of
computation), and our advantage over the Fokker-Planck treatment is
the inclusion of binaries, and better stellar evolution.  Our
principle disadvantage, compared with the $N$-body modelling, is the
need for simplifying assumptions in the treatment of dynamical
encounters of all kinds; compared with the work of \citet{Dr1995}, our
main shortcoming is our very limited ability to assess the uniqueness
of the model we eventually arrive at.

In the following section we summarise some global data about NGC6397
and its close relative, M4.  We also describe the data which we use in
assessing our models, so that they match various aspects of the
cluster at the present day.  Next we describe our best models, and
include an assessment of the reasons why the two clusters have such
dissimilar surface brightness profiles.  In the final section we sum
up and discuss aspects of our initial conditions more generally.

\section{Data on NGC 6397}\label{sec:data}

Global data relevant to our study of NGC6397 is listed in Table 1,
along with data for M4, for comparison.  In most cases the sources
given are not the original sources; nevertheless some of the sources
include discussion of the literature and of the data selected.  { We
became aware of the age determination given in this table only after
our research was well under way, and compare our model at an age of 12Gyr,
as we did for M4.}

\begin{table}
\begin{center}
\caption{Data on NGC6397 and M4 (NGC6121)}
\begin{tabular}{lll}
&NGC6397&M4\\
\hline
$R_{GC}^1$ (kpc)& 6.0&5.9\\ 
${R_\oplus}^2$ (kpc)&$2.55^5$&$1.72^3$\\
$A_V$&$0.56^4$&$1.33^3$\\
$(m-M)_V$&$12.03^4$&$11.18^3$\\
$[Fe/H]^1$& -1.95&-1.20\\
$M_V^1$&-6.63&-7.20\\
$M (M_\odot)$&$^66.6\times10^4$&$^36.25\times10^4$\\
$r_c$ (arcmin)$^1$&0.05&0.83\\
$r_h$ (arcmin)$^1$&2.33&3.65\\
$r_t$ (arcmin)$^1$& 15.81&32.49\\
{ Age (Gyr)$^7$}&{ $11.47_{-0.47}^{+0.47}$}&{ $12.1_{-1.8}$}\\
\end{tabular}
\label{tab:data}
\end{center}
Sources and notes: $^1$ \citet{Ha1996}; $^2$ Distance from
earth; $^3$ \citet{Ri2004}; $^4$ \citet{Ri2008}; $^5$ from $m-M$; $^6$
\citet{Dr1995}; $^7$ \citet{Ha2004, Ha2007}
  \end{table}

For the surface brightness our main source, as with M4, has been
\citet{Tr1995}, but we have also considered the HST profile of the
central parts of the cluster of \citet{NG2006}.  These are in
reasonable agreement where they overlap, except at the very centre,
where the HST profile is slightly brighter.  To some extent the
agreement is artificial, as Noyola \& Gebhardt calibrated their
surface brightness determinations against the older data.

The velocity dispersion profile given by \citet{MM1991} is based on
radial velocities of 127 stars.  We have also been kindly given access
to a catalogue of unpublished radial velocities of over 1400 stars
observed by the Padova group (G. Piotto, pers.comm.).  The reduction
of this data is not yet complete, however, and we have made no use of
them here.

While there is no shortage of luminosity functions in the literature
for NGC 6397, relatively few of these are properly normalised,
i.e. give counts in a stated area of sky.  All too often only the
shape of the luminosity function has been published.  Our primary
sources of absolute luminosity functions are restricted to
\begin{enumerate}
  \item \citet{Pa1995}, who give counts in the F814W band in a field about 4.6 arcmin
  from the cluster centre; and
\item \citet{Mo1996}, who gave similar results for a field at about 10 arcmin.
\end{enumerate}
{ For purposes of comparison with our models, which give $I$-band
magnitudes, we have ignored the difference between $m_I$ and
$m_{814}$.  From the models tabulated in 
\citet{Ba1997} we estimate that the  effect of this on the
luminosity function  is less than 5\%.}

Finally we mention the  observed binary fraction.  From the
binary sequence in the colour-magnitude diagram, \citet{Da2008a}
conclude that this decreases from about 5\% at the centre to about 1\%
at 5 arcmin.  On the other hand A. Milone (pers. comm.) estimates that
the outer
binary fraction is about 3-4\%, rising to about 9\% in the inner 15
 arcsec.  
 Clearly, there is some uncertainty here.  For later
discussion, these results are presented, along with similar
information for M4, in Table \ref{tab:fb}.

\begin{table}
\begin{center}
\caption{Approximate observed binary fractions in  NGC6397 and M4}
\begin{tabular}{ccccl}
\hline
M4&M4&NGC 6397 &NGC 6397&Source\\
centre&off-centre&centre& off-centre&\\
\hline
10\%&--&9.2\%$\pm$3.0&3.7\%$\pm$0.8&$a$\\
2\%&1\%&5.1\%&1.2\%&$b$
\end{tabular}
\end{center}
Sources: $a$ \citet{Mi2008} and Milone, pers.comm.
; $b$ \citet[M4]{Ri2004}, \citet[NGC 6397]{Da2008a}
\label{tab:fb}
  \end{table}

\section{Models of NGC 6397}

\subsection{ The search for a model}\label{sec:params}

Our general approach is to follow a cycle in which we (a) specify a
model, (b) follow its evolution with the
Monte Carlo code, (c) compare with observational data on surface
brightness and velocity dispersion profiles, and luminosity functions
and (d) alter the initial model to improve the fit.  { Though complete models
of NGC6397 now take only about a day on a good PC, thanks to recent
software improvements, }
 we still try to
accelerate the first few cycles by using scaled-up results from
smaller models \citep{HG2008}.  From the same source we list in {
  the first column of} Table
\ref{tab:ics} the free parameters which we have to adjust in step
(a) of each cycle{, except that the total mass $M$ is determined by
  the other parameters}.

\begin{table}
\centering
\begin{minipage}{140mm}
\caption{Initial parameters for models of NGC6397, and ranges sampled}
\begin{tabular}{lllll}
  Number of stars, $N$&{ $4\times10^4$}&{ $10^5$}&{ $5\times10^5$}\\
Mass,		$M$\\
Tidal radius, $r_t$ (pc) &{ [66.2:80.9]}&{ [56.0:61.9]}&{ [40]}\\
Half-mass radius, $r_h$&{ $r_t$/[60:100]}&{ $r_t$/[95:105]}&{
  $r_t$/[60:105]}\\
Binary fraction,	$f_b$&{ [0.03:0.09]}&{ [0.06:0.09]}&{
  [0.03:0.15]}\\
Slope of the lower IMF, $\alpha$&{ [0.7:1.3]}&{ [0.7:1.0]}&{ [0.3:1.1]}\\
\hline
\end{tabular}
\label{tab:ics}
\end{minipage}
  \end{table}


Since the present-day parameters of NGC 6397 are quite similar to
those of M4 (Table \ref{tab:data}), we started close to the parameters
which we had reached for M4 in \citet{HG2008}.    Our scaled-down models
with $N = 40 000$ and $100 000$ stars coarsely covered the ranges stated in the
second and third columns of Table \ref{tab:ics}.  This information requires some
explanation.  First, we have not stated the corresponding 
total mass $M$; as we have said, this is determined from the other stated parameters
($N, f_b$ and $\alpha$) for our choices of the initial mass functions
of single stars and binaries, and the distribution of binary component
ratios.  Second, the ranges of tidal radius $r_t$ look disjoint, but
scaling by $N$ also requires scaling lengths, in order to keep the
relaxation time fixed (see \citet{HG2008}).  Also, we often
internally used equivalent values for $N= 10^4$, which explains the
odd choices listed here.   Third, we specified our
initial half-mass radius in terms of the ratio $r_t/r_h$, which
explains the way in which we have expressed $r_h$.  Finally, each
model can be scaled (approximately) to a different value of $N$, and
we have not recorded the ranges with which we experimented.

By comparing these models (by eye) with the observational data we have
mentioned, we located a promising region in this parameter space.
This was then explored with a full-sized model over parameter ranges
given in the fourth column of the Table.  This exploration was less
systematic, however, and not all combinations of parameters were
considered.  Also, we extended somewhat the ranges of $f_b$ (the
binary fraction) and $\alpha$ (the power law index of the lower
initial mass function), as indicated. 

Before we describe one of our best models in detail, we record here
the effect of departing significantly from these apparently
near-optimal values of the parameters.  
\begin{enumerate}
  \item $N$: this can be explored (approximately) by scaling, which
  leads to the following conclusions.  A 25\% increase in $N$ leads to
  an overall increase in the surface brightness by about 0.4
  magnitudes, an overall increase in the velocity dispersion profile
  (i.e the root mean square line-of-sight velocity) by about 15\%, and
  an increase in the luminosity functions by about 30\%.  Note that
  the scaling we employ preserves the relaxation time, and so lengths
  are scaled as well as $N$ itself \citep{HG2008}.
\item $r_t$: an increase by 10\% causes little change  in the  surface
brightness profile except above about $0.5r_t$, where it causes an
approximately proportional spatial extension of the profile.  An
increase by 10\% causes approximately the same percentage increase in
the velocity dispersion, and large increases in the luminosity
function: by factors of approximately 2 and 3 in the inner and outer
fields, respectively.  
\item $r_h$: A 10\% increase causes a slightly less concentrated
  surface brightness profile, changing the surface brightness by only
  a few tenths of a magnitude.  It produces about a 3\% decrease in
  the velocity dispersion profile, and decreases in the luminosity
  functions by about 20\%.
\item $f_b$:  A doubling of the primordial binary fraction (from 0.03
  to 0.06) has no discernible effect on the surface brightness profile
  or the luminosity functions, and only a
  small (of order 2\%) increase in the velocity dispersion.
\item $\alpha$: An increase in the power-law index of the lower
  initial mass function by 0.3 results in a decrease in the surface
  brightness by about 0.2--0.3 magnitudes, and a decrease in the velocity
  dispersion profile by about 8\%.  The luminosity function is almost
  unchanged at the bright end, but increases by about 25\% at the
  faint end.
\end{enumerate}

In considering these remarks, it must be remembered that they refer to
the range of radii and magnitudes for which we had observational data,
which are displayed in Figs.\ref{fig:sbp}--\ref{fig:lf2}.  This helps
to explain why an increase in the tidal radius can have a much larger
effect on the luminosity function than on the surface brightness,
because these refer essentially to  fainter and brighter stars,
respectively.  An increase in tidal radius causes less loss of
low-mass stars, and so it can be understood qualitatively how the
stated changes in the various observational parameters arise, when
$r_t$ is increased.

The above remarks also show that some parameters are much more easily
determined than others.  The binary fraction is best determined by
comparison with the observed binary fraction (Table \ref{tab:fb}), and
is very poorly constrained by the surface brightness and velocity
dispersion profiles, and the luminosity function.  On the other hand
the effect of $r_t$ on the luminosity function is so great that it is
well determined (for given values of the other parameters).  Of the 
remaining three parameters we judge that $\alpha$ is the least well
determined. 

Eventually this exploration 
led to a small number of satisfactory models, one of which 
 is summarised 
in Table \ref{tab:ic-values}.  There we show for comparison the
corresponding results from the most satisfactory model of M4, and a
few interesting numerical results. { The binary fraction  needs
some explanation, as it appears there has been much more depletion in
the model of NGC6397 than in the model of M4.  The reason for this is
that the M4 model used a uniform initial distribution of semi-major
axis, while study of the radial velocity binaries in M4 detected by
\citet{So2008} has suggested to us that a better fit is provided by
models with the semi-major axis distribution given by the procedure
described by \citet{Kr1995}.  This gives a higher proportion of soft
binaries, and hence enhanced destruction.}

Since carrying out this search for a model, we have also codified an
extensive grid of small scale models in such a way that initial values
can be found as a function of corresponding values at the present
day.  This does not, however, fully solve the problem of determining
initial conditions, as present-day values of these parameters are not
often known reliably, and a fit to values drawn from the literature
does not guarantee a fit to an entire surface brightness profile,
etc.  Nevertheless, this procedure may in future provide plausible
starting values for a more refined search.

\begin{table*}
\begin{center}
  \begin{minipage}{130mm}

\caption{Initial and final conditions for models of NGC6397 and M4}
\begin{tabular}{lllll}
\hline
&NGC6397: 0Gyr&NGC6397: 12Gyr&M4: 0Gyr&M4: 12Gyr\\
\hline
Mass
($M_\odot$)&$3.65\times10^5$&$6.03\times10^4$&$3.40\times10^5$&$4.61\times10^4$\\
Luminosity
($L_\odot$)&$5.1\times10^6$&$3.40\times10^4$&$6.1\times10^6$&$2.55\times10^4$\\
Binary fraction 	&0.090&0.044&0.07&0.057\\
Slope of the lower mass
function&1.1&0.48&0.9&0.03\\
Mass of white dwarfs
($M_\odot$)&0&$2.74\times10^4$&0&$1.81\times10^4$\\
Mass of neutron stars ($M_\odot$)&0&$3.14\times10^3$&0&$3.24\times10^3$\\
Tidal radius (pc)&40.0&22.0&35.0&18.0\\
Half-mass radius (pc)&0.40&3.22&0.58&2.89\\
{ Core radius$^1$ (pc)}&0.22&$0.041\pm0.012$&0.31&$0.043\pm0.017$\\
\hline
\end{tabular}
\label{tab:ic-values}

{$^1$The definition is given in the caption of Fig.\ref{fig:radii}.
For 12Gyr the values stated are the mean and 1-$\sigma$ variations
over the period 11--12Gyr.  For 0Gyr the value uses the same formula,
but analytic values for the central density and velocity dispersion.}
  \end{minipage}
\end{center}
  \end{table*}

\subsection{The surface brightness profile}\label{sec:sbp}

  \begin{figure}
{\includegraphics[height=12cm,angle=0,width=9cm]{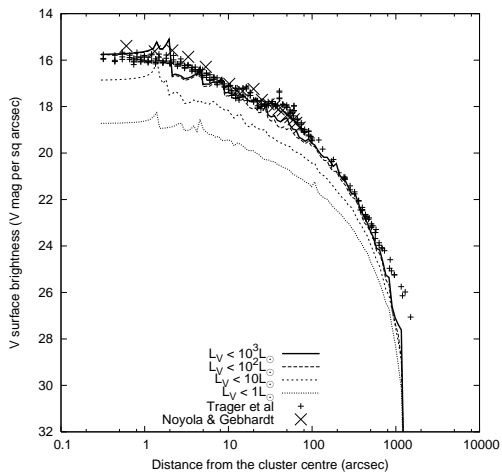}}
    \caption{Surface brightness (in units of V magnitudes per square
    arc second) as a function of projected radius (arc seconds), in
    the model and in the observational data from
    \citet{Tr1995} and \citet{NG2006}.  { Only the data classed by
    Trager et al as of highest weight are included.  }Several model profiles are
    given, corresponding to various limits on the brightest star
    included.  The profiles for $10^3$ and $10^2L_\odot$ differ little.}
\label{fig:sbp}
  \end{figure}

Fig.\ref{fig:sbp} shows how the surface brightness profile of our
model compares with the ground-based data of \citet{Tr1995} and the
HST data of \citet{NG2006}.  { Except at small radii,} the model is slightly faint compared with
the former, which in turn is fainter than the latter.  The model is
especially faint, and too small, at large radii, but we have explained
in \citet{GHH2008} that this is an artifact of our treatment of the
tide. 

{ Next}  we should explain the reason for the peaks in the model surface
brightness profile, which are particularly obvious at small radii.
The Monte Carlo code gives the radius and V-magnitude of each star,
and, in order to construct the surface brightness, we represent each star
by a sphere of the corresponding radius and magnitude.  The projected surface
brightness of a sphere is infinite at its edge, and the peaks occur
at projected radii close to the radii of stars, especially bright
ones.  From an observational point of view also the surface brightness
has very large fluctuations (at the locations of the stars), and these
are reduced by averaging over annuli, exclusion of bright stars, etc.

The effect of excluding bright stars in the model is shown in
Fig.\ref{fig:sbp} by four profiles in which successively fainter
bright stars are excluded, down to roughly the turnoff.  As expected,
the overall level and fluctuations are reduced, but these plots do
little to clarify the nature of the core in the model.  When the
fluctuations are sufficiently suppressed, the profile is severely
biased.

\begin{figure}
{\includegraphics[height=12cm,angle=0,width=9cm]{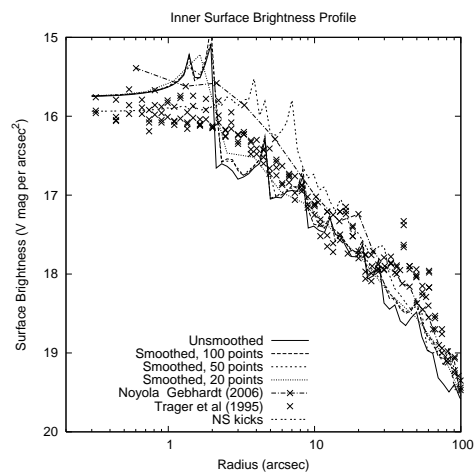}}
    \caption{Surface brightness (in units of V magnitudes per square
    arc second) as a function of projected radius (arc seconds), in
    the inner part of the cluster.  For the model, various smoothed
    profiles are shown, and the key states the number of intervals,
    evenly spaced in $\log R$, over the interval plotted in
    Fig.\ref{fig:sbp}.   The bin sizes in $\log R$ are 0.037, 0.074
    and 0.185, respectively.  The observational data are as in
    Fig.\ref{fig:sbp}.  The curve labelled ``NS kicks'' is discussed
    in Sec.\ref{sec:nskicks}.}
\label{fig:smooth}
  \end{figure}

Another way of suppressing fluctuations, i.e. smoothing in radius, is
illustrated in Fig.\ref{fig:smooth}, which focuses on the inner part
of the cluster.  The most noticeable difference between the various
profiles is, once again, the fact that the profile of \citet{NG2006}
is somewhat brighter { at most radii}.

\citet{NG2006} give a short but interesting review of the literature
on the surface brightness profile of NGC6397, and point out that,
despite its status as a collapsed core cluster, various authors have
given core radii of order a few seconds of arc.  Part of the reason
for the difficulty of classifying the cluster may be gleaned from the
$V$ surface brightness map given by \citet{Au1990}.  The area within 7
arcsec of the centre is dominated by a roughly linear group of blue
stragglers, and it is not clear how this influences a radially
averaged profile.

{
The difficulties with the observational material are not confined to
the centre, however.  It is evident from Fig.\ref{fig:smooth} that,
even though this includes only material regarded by Trager et al as of
highest weight, there are regions within the range 10--100 arcsec
where their data is not self-consistent.  In particular there is a ``bump''
around 40 arcsec (or possibly a trough at smaller radii) which could
be fitted only by a highly contrived model.  It is important to note also
that the data of Noyola \& Gebhardt is calibrated with the aid of the 
older data, by matching the integrated light of the newer data to the
integral of a smooth fit to the older data.  }


We shall see in Sec.\ref{sec:fluctuations} that
the profile of the model is subject to considerable fluctuations, both in time and
for different realisations of the same model.   { Therefore the
  question of whether the model is in agreement with the observational
  profile, even if the latter were completely reliable, has to be
formulated statistically; in other words, can the observational
profile be rejected as a member of the ensemble of profiles provided
by the model, or not?  We do not answer this question in this paper,
but in this spirit we do provide a discussion in terms of the central
surface brightness in Sec.\ref{sec:fluctuations}.

There is another way of approaching the question of whether the
surface brightness profile of the model matches that of NGC6397, at
least qualitatively, and
that is to ask whether our model would be
taken for a collapsed-core cluster.}   Even the most
smoothed model profile in Fig.\ref{fig:smooth} is not monotonic, but { all suggest that
this is a model of a cluster which would be classified as having a
cusped profile rather than a cored one.}
Note that, while we have not
attempted any quantitative measure of this statement,  
to some extent we have here
followed the lead of \citet{Tr1995}, who classified the observed
profiles of star clusters { apparently} without any numerical criterion for doing so.


\subsection{Velocity dispersion profile}

\begin{figure}
{\includegraphics[height=12cm,angle=0,width=9cm]{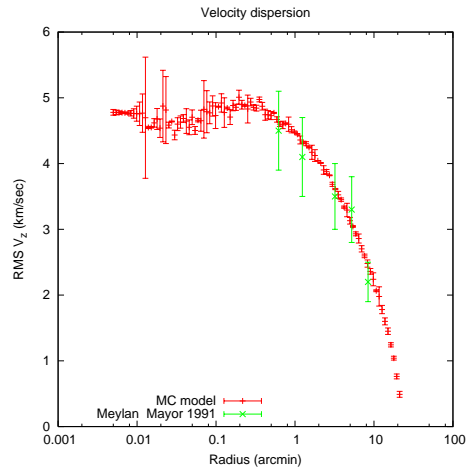}}
    \caption{Line-of-sight velocity dispersion (km/sec)  as a function
    of projected radius (arc min)  for the Monte Carlo model, and from
    the observational data of \citet{MM1991}.
}
\label{fig:vdp}
  \end{figure}

The purpose of Fig.\ref{fig:vdp} is to illustrate the satisfactory fit
between the line-of-sight velocity dispersion of the model and the
observational data of \citet{MM1991}.  Only stars brighter than $M_V =
7$ are included, as the observations are confined to giants and
subgiants.  { In the model, binaries are represented by the velocity of
their centre of mass; internal motions are neglected.}
The slight  depression in the
 velocity dispersion near the centre may be attributable to mass segregation;
similar qualitative features occur in the more idealised models of
\citet{CW1990}.  {  No such  central inversion in the velocity
dispersion was noticeable in our models of M4 \citep{HG2008}, however,
 and this may be associated with the fact that  the radius of the central
 concentration of heavy remnants in our model of NGC 6397 is somewhat
 smaller than in our model of M4.
(Note that, in both models, neutron stars and white dwarfs receive no
 natal kick.) 

}

\subsection{Luminosity functions}

\begin{figure}
{\includegraphics[height=12cm,angle=0,width=9cm]{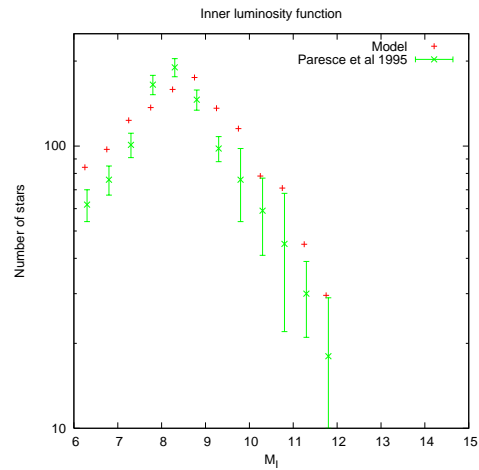}}
    \caption{I-band luminosity function in a field at a radius of
    about 4.6 arcmin \citep{Pa1995} compared with the model.  Strictly
    the observational magnitude is $M_{814}$.
}
\label{fig:lf1}
  \end{figure}

\begin{figure}
{\includegraphics[height=12cm,angle=0,width=9cm]{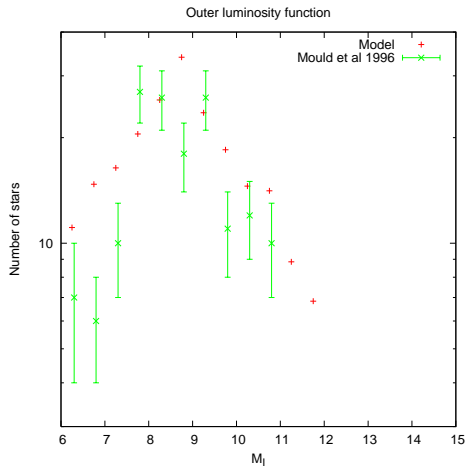}}
    \caption{I-band luminosity function in a field at a radius of
    about 10 arcmin \citep{Mo1996} compared with the model.
}
\label{fig:lf2}
  \end{figure}

Main sequence luminosity functions are displayed, for the model and two observational data sets, in
Figs.\ref{fig:lf1} and \ref{fig:lf2}. {  For the model, we counted all
objects in the colour-magnitude diagram above a line just below the
main sequence.  These luminosity functions therefore include
unresolved binaries (in the same way that these were not removed from
the observational data with which we compare).  

  While the surface brightness profile (Fig.\ref{fig:sbp}) gives the
impression that the model is slightly faint at most radii, the model luminosity
functions generally seem somewhat too high, are not steep enough at
the bright end, and peak at somewhat fainter magnitudes (compared with
the observational data.)  These discrepancies seem quite recalcitrant; considerable experimentation
with the initial conditions of the models has been unable to improve
this aspect of the comparison with observation, though it has to be
noted  that, as far as the IMF is concerned, we have limited ourselves
to varying  the slope of the lower
mass function.  (The break in the { slope of the assumed} mass function  occurs at $0.5M_\odot$,
corresponding to about $M_I = 6.9$.)  
{ In any event} 
it must be remembered that
the { observational data} are affected by incompleteness, especially at the faint end in
the inner field.  Furthermore, \citet{Mo1996} show that independent
counts in the inner field at the bright end may differ by as much as
30\%.}  { Finally, while the model luminosity functions correspond to
the stated radius, the observational luminosity functions are
determined in fields which cover a range of radii, and stellar
densities.}  { Our experience is that the overall fit to the
  luminosity function depends sensitively on some of the initial
  parameters of our models (Sec.\ref{sec:params}), and so we regard  the
  agreement achieved with our model as not unsatisfactory.}

\subsection{Binary fraction}

\begin{figure}
{\includegraphics[height=12cm,angle=0,width=9cm]{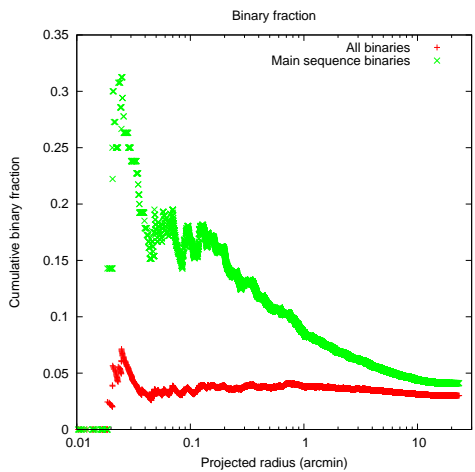}}
    \caption{Projected cumulative binary fraction for all objects and
    for objects on or above the main sequence.  The ordinate is the
    binary fraction for all objects with projected radius less than
    the value on the abscissa.  In computing this fraction for objects
    on or above the main sequence, both numerator and denominator
    exclude white dwarfs and other degenerate objects.  }
\label{fig:binp}
  \end{figure}

We have already mentioned the difficulty in assessing observationally
the binary fraction in this cluster.  As for the model, the result
depends on whether we include all objects, including white dwarfs, or
only objects on or above the main sequence (Fig. \ref{fig:binp}).  For
the latter, the binary fraction drops rapidly with increasing radius,
{ roughly in qualitative agreement with the observational results (Table
\ref{tab:fb}).  We assume that it would be equally easy to produce a model
in rough {\sl quantitative} agreement, simply by
reducing the initial binary fraction in the model.  Our reason for
presenting results using an excessive binary fraction is that we
wanted to address the question of whether it is the binary fraction
which determines whether a post-collapse cluster exhibits a cusped or
cored profile (Sec.\ref{sec:bf}).  We already see that this is not the
case: { here we have a model with a non-King surface brightness
  profile but an excessive binary fraction.}}

\subsection{Neutron star kicks}\label{sec:nskicks}

{
One of our aims in this paper (Sec.\ref{sec:comparison})
is to compare our modelling of the two
clusters M4 and NGC6397.  Therefore we have in general adopted similar
astrophysical assumptions for NGC6397 as for M4 in \citet{HG2008}.  (One exception is the initial distribution
of binary periods, but it has already been mentioned that this has
little discernible effect on the fit to the observational data shown
in Figs.\ref{fig:sbp}--\ref{fig:lf2}.)  In our modelling of M4 we gave
no natal kicks to neutron stars, and have adopted the same assumption
here.  Though there are grounds for considering that neutron stars may
indeed be born with low kick velocities (e.g. \citealt{Pf2002}), we
here consider one model with a more conventional kick distribution,
with a one-dimensional velocity dispersion of 190km/s.

We did not attempt to iterate on the initial parameters to optimise
the fit with the observational data, but used those given in
Tab.\ref{tab:ic-values}.  In the resulting model the time of core
collapse was delayed by about 4Gyr to 9Gyr, and by 12Gyr the mass of
neutron stars was smaller by a factor of about 20.  Nevertheless the
model 
exhibited a velocity-dispersion profile and luminosity functions very
similar to those of the model with no neutron star kicks
(Figs.\ref{fig:vdp}--\ref{fig:lf2}), including the imperfections of
the latter.  The main difference was in the surface brightness
profile, the central part of which is displayed in
Fig.\ref{fig:smooth}.  This generally matches the data of
\citet{NG2006} better than our standard model, but not in the
innermost 1 or 2 arcsec, and it has the character of a King-like
profile with a small core.  Therefore, while a model with neutron star
kicks may be a better basis on which to optimise over the initial
parameters, it does not of itself provide an explanation for the
non-King profile of NGC6397.  }

\section{Comparison of M4 and NGC 6397}\label{sec:comparison}

\begin{figure}
{\includegraphics[height=12cm,angle=0,width=9cm]{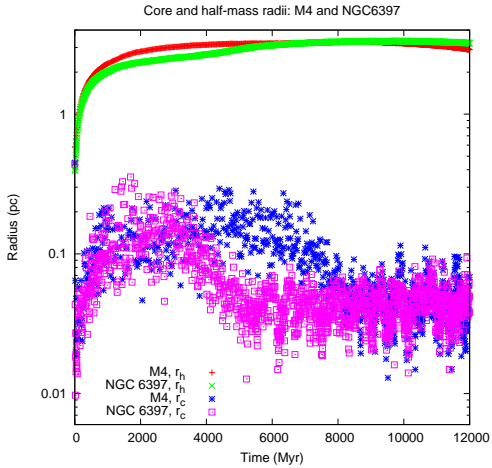}}
    \caption{Evolution of the half-mass and core radii of our models
    of NGC 6397 and M4.  Note that  what is plotted here is the {\sl
  dynamical} core radius, which is computed by the formula
$
r_c^2 = {3\langle v^2\rangle}/({4\pi\rho_0}),$
where the mass-weighted mean three-dimensional velocity $\langle
v^2\rangle$ and the central density $\rho_0$ are calculated for the
innermost 20 stars.
}
\label{fig:radii}
  \end{figure}

What is striking about Table \ref{tab:ic-values} is the similarity of
the initial conditions for these two clusters, and yet one (M4) is
described traditionally as a `King-model' cluster, while the other
(NGC 6397) is classified as `post-core-collapse' \citep{Tr1993}.  As
we have seen (Sec.\ref{sec:sbp}) the { observed} surface brightness profile of NGC
6397 is not entirely unambiguous in this respect, while the
similarities between the two clusters were further strengthened by
\citet{HG2008}, whose model of M4 suggested that this cluster
experienced core collapse at an age of about 8Gyr.  (Our model of
NGC6397 is also post-collapse, the collapse of the core having ended
at a somewhat smaller age of 5Gyr; see Fig.\ref{fig:radii}.)

\begin{figure}
{\includegraphics[height=12cm,angle=0,width=9cm]{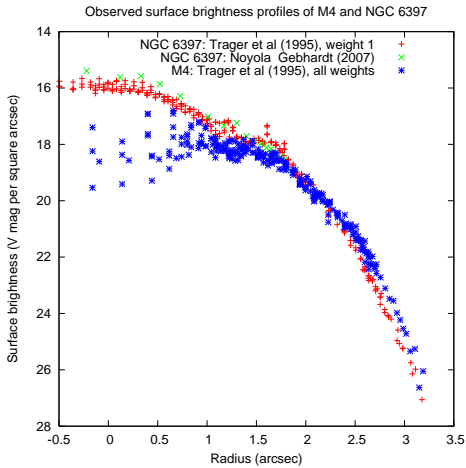}}
    \caption{Observational surface brightness profiles of M4 and NGC
      6397.  For NGC 6397 only the data of highest weight (in the
    estimation of \citet{Tr1995}) are included.
}
\label{fig:obs-sbp}
  \end{figure}

\begin{figure}
{\includegraphics[height=12cm,angle=0,width=9cm]{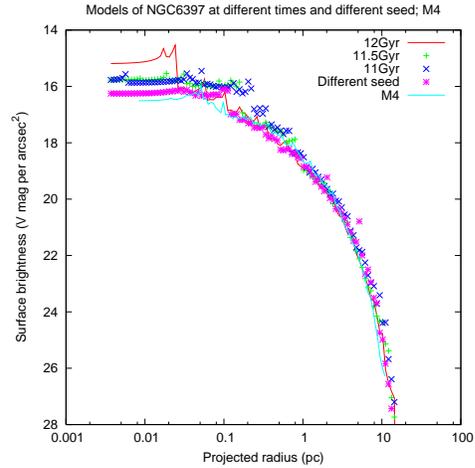}}
    \caption{ Surface brightness profiles of a model of NGC
      6397 at ages of 11, 11.5 and 12 Gyr, and a second model with a
      different initial seed at 12 Gyr.  { Also shown is a model of M4
      \citep{HG2008}.  Because the comparison involves different
      clusters, the surface brightness of the models has not been
      corrected for extinction, and the units of the radius are parsecs.}
}
\label{fig:different-times}
  \end{figure}

The distinction between the two clusters on the basis of the observed surface
brightness profile is, nevertheless, striking
(Fig.\ref{fig:obs-sbp}).  Only slightly less striking is a similar
comparison of the model surface brightness profiles
(Fig.\ref{fig:different-times}).  The availability of these models allows us to
investigate, and even experiment with, the possible reasons for the
difference between these clusters.  Note that there are two questions:
why the two clusters differ, and why the {\sl models} of the two
clusters differ.  We approach the second question in this section.  Mechanisms
which can account for differences in the models may help to account
for the differences in the actual clusters, but there are additional
issues which may come into play in reality, and some of these are
addressed in the final section of the paper.

\subsection{Fluctuations}\label{sec:fluctuations}

One possible explanation for the difference between the models of the two clusters
is fluctuations, and this hypothesis falls into at least two  subclasses.
First, we may suppose that, because of the motions of the stars, or
possibly the occurrence of core oscillations \citep{SB1983}, the
profiles at different times will naturally differ.  A second
possibility is that different realisations of the same initial
conditions could give rise to significantly different surface
brightness profiles.  We consider these two possibilities in turn.

\subsubsection{Fluctuations in time}\label{sec:csb}

Fig.\ref{fig:different-times} shows surface brightness profiles of a
Monte Carlo model of NGC 6397 at three successive times
separated by 500Myr. { While one  might expect to see  a trend for the central surface
brightness to decrease, this is obscured (in the present example) by
fluctuations, which actually lead to an increase in the central
surface brightness { with time}.}  Though we have plotted only three profiles from
this model, they give the impression that fluctuations could be sufficient to account fully
for the difference between the surface brightness profiles of our
models of the two clusters M4 and NGC 6397:
Fig.\ref{fig:different-times} also displays the surface brightness
profile of our model of M4 \citep{HG2008}, plotted to the same
physical length scale, and with all profiles uncorrected for extinction.
 
\begin{figure}
{\includegraphics[height=12cm,angle=0,width=9cm]{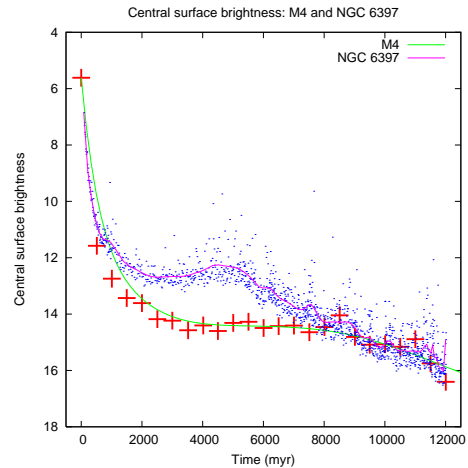}}
    \caption{Evolution of the central surface brightness (in units of
    V magnitudes per square arcsec) of our models
    of NGC 6397 and M4.  The smoothed curves are distorted by the large
    values at $t=0$, but elsewhere they are a satisfactory guide.  The
    data are not available at the same frequency in the two models,
    but since the M4 model has been published, we did not attempt to
    rerun it.
}
\label{fig:csb}
  \end{figure}

While the small number of profiles plotted in
Fig.\ref{fig:different-times} may be { no more than} suggestive, a more complete
picture is provided by  the evolution of the central surface brightness
(Fig.\ref{fig:csb}). 
{ Though the mean central surface brightness at the present day may be
slightly larger for NGC6397 than for M4, any such difference is
obscured by fluctuations.  Indeed}  the fluctuations
are  large enough that the centre of NGC6397 is often
dimmer than the centre of M4.    Guided by
Fig.\ref{fig:different-times} this would imply also that it often 
has the {\sl less}  compact core.  These results imply that it is no
more than an accident that, at the present day, NGC6397 presents a
cusped profile, while the profile of M4 is classified as cored.

\begin{figure}
{\includegraphics[height=12cm,angle=0,width=9cm]{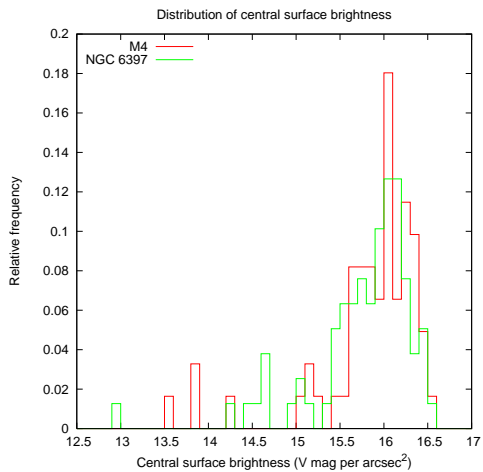}}
    \caption{ Histogram of the values of the central surface brightness
    in the models of M4 and NGC6397 between 11.6 and 12 Gyr.
}
\label{fig:csb-hist}
  \end{figure}

{ { As explained}  in the caption to
  Fig.\ref{fig:csb}, we do not have data on the central surface
  brightness of our M4 model at the same frequency as for NGC 6397.
  We have, however, run a similar model of M4, and it confirms
  (Fig.\ref{fig:csb-hist}) that
  the distribution of central surface brightness is almost
  indistinguishable between the models of the two clusters, at least
  at the present day.  (Fig.\ref{fig:csb} shows that
the model of  NGC6397 will have been brighter at earlier ages.)}

{ The central surface brightness of our model at 12Gyr is 15.2
(uncorrected for extinction), and it can be shown from the data used
  to plot Fig.\ref{fig:csb-hist} that this value is exceeded (i.e. the
  centre is brighter) about 14\% of the time.  While this may seem to
  imply that a surface brightness profile like that of NGC6397 is
  slightly unusual, it must be remembered that NGC6397 is itself
  actually slightly dimmer at the centre than our model, and so the
  proportion of models exhibiting a non-King profile may well exceed
  14\%.  Furthermore, this percentage depends somewhat sensitively on
  the overall brightness of the model, which we know is slightly dim
  (Sec.\ref{sec:sbp}), and on the dynamics of the degenerate
  population (Sec.\ref{sec:nskicks}).}

\subsubsection{Fluctuations between different realisations}

What we have in mind here is not simply the fact that the masses and
positions of stars  will be different in different realisations,
i.e. runs carried out with different seeds for the random number
generator.  More interesting is the possibility that the behaviour of
a population of rare objects, present in sufficiently small numbers to
be subject to large statistical fluctuations, could have a dramatic
effect on the entire system.  An example of this
is Hurley's important discovery of the
effect of binary black holes \citep{Hu2007}.  He considered two different
realisations of the same $N$-body model, which had $N = 10^5$ single stars
initially, with a mass spectrum and stellar evolution.  In one of
these realisations, a stellar-mass black hole binary happened to form
dynamically, and it had a dramatic effect on the evolution of the core
radius.  Towards the end of the simulation, the core radius was about
twice the value in the other realisation, where, by chance, no black hole binary
was formed.  Though black holes are expelled in our models at about
the time of core collapse (Fig.\ref{fig:radii}), { and though our
  models are richer than Hurley's,} this process is
  indicative of the kind of fluctuations we have in mind.

Two realisations of the same model are shown in
Fig.\ref{fig:different-times}, and they
illustrate the fact that different realisations lead to { more
 modest variations in the surface brightness
profile than the variations (due to the presence or absence of a
binary black hole) found  by Hurley.  Indeed, further study shows, not
surprisingly, that
the fluctuations between models with different seeds are comparable
with those in time within one model (Figs.\ref{fig:different-times},\ref{fig:csb-hist}).}

As a cautionary remark, we should point out that the behaviour of
fluctuations in a Monte Carlo model could, in principle, differ from
their behaviour in an $N$-body model, though we have no reason to
suspect this.  The point is that, though we have been to considerable
pains to ensure that the evolution of the total mass and various other
parameters behaves similarly in our Monte Carlo models and in
comparable $N$-body models \citep{GHH2008}, we did not check that
fluctuations also behave similarly.  In principle, each time some new
feature is studied with a Monte Carlo model, it should be validated
with a check against a model with fewer simplifying assumptions.  {
  In a forthcoming paper we shall present a comparison between our
  Monte Carlo model of NGC6397 and the evolution of an $N$-body
  realisation of the same model.}

\subsection{Different initial and boundary conditions}

\subsubsection{Different initial concentration}\label{sec:init-conc}

\begin{figure}
{\includegraphics[height=12cm,angle=0,width=9cm]{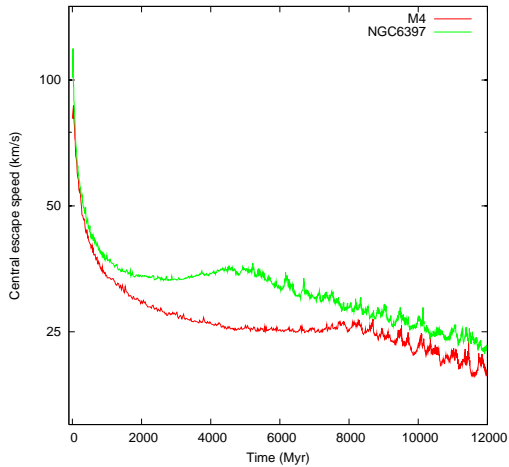}}
    \caption{Evolution of the central escape speed in our models
    of NGC 6397 and M4.  The ordinate is logarithmic.
}
\label{fig:escape-speed}
  \end{figure}

We already drew attention \citep{HG2008} to the high initial
concentration of our model of M4, and our initial model of NGC6397 is
even more compact.  (The initial half-mass radii are given in Table
\ref{tab:ic-values}.) { It is tempting to suppose naively that the
effect of this is a more concentrated configuration at an age of 12
Gyr, { and that this alone is the explanation of the difference between
the surface brightness profiles of the two clusters.  We have already
mentioned, however, that the fluctuations in the central surface
brightness are much larger than any
systematic difference; it is only earlier in
the evolution of these clusters that NGC6397 was systematically
brighter  than M4.    The evolution of the half-mass radius is also generally
very similar for both clusters (Fig.\ref{fig:radii}), even though
initially the half-mass radius of our model of M4 was almost 50\%
larger than that of the model of NGC6397.

One can see systematic differences in the two clusters by study of the
  central escape speed, which  in our model of NGC 6397
  always exceeds that in our model of M4
  (Fig.\ref{fig:escape-speed}).  The reason why these differences are
  more apparent  is that
  fluctuations in the core have a much smaller effect on the central
  escape speed than on the central surface brightness, as so much of
  the central potential is determined by the total mass and half-mass
  radius of the cluster.}

The
higher initial concentration should lead to enhanced destruction of
binaries, but the overall binary fraction is low enough that this
difference has no discernible effect on the surface brightness profile.}

\subsubsection{Other factors}

While we have attributed the { systematic} differences in the two curves in
Fig.\ref{fig:csb} to the initial concentration, there are several {
  other}
differences between the initial conditions of the models depicted
(Tab.\ref{tab:ic-values}).  But tests have shown that, if we vary each
initial condition one at a time, all have a minor effect on the
evolution of the central surface brightness, except for the initial concentration.  In
particular, changing the binary fraction 
has little effect compared with the difference of
initial half-mass radius.


\section{Conclusions and Discussion}

\subsection{Conclusions}

We have constructed a dynamical evolutionary model for the Galactic
globular cluster NGC 6397.  The model is based on a Monte Carlo
treatment of dynamical interactions (2-, 3- and 4-body), and synthetic
treatments of the evolution of single and binary stars.  By varying a
number of initial parameters ($N$, tidal and half-mass radii $r_t$ and
$r_h$, binary fraction, slope of the lower IMF; see Tables
\ref{tab:ics},{ \ref{tab:ic-values}})
we have constructed a model which, after 12Gyr of evolution, resembles
NGC 6397 in terms of the surface brightness profile, the velocity
dispersion profile, and the luminosity function at two radii.

{ Like the clusters themselves, our model for NGC 6397 has a higher
central surface brightness than the model for M4, which is associated
with the fact 
that, while M4 has a surface brightness profile resembling a King
model, NGC 6397 has a more compact central profile which has led to
its classification as a non-King cluster.  Though this is associated
in the literature with the distinction between pre- and post-collapse
clusters, we find that both clusters are in the post-collapse phase of
evolution.  {\sl Fluctuations are the dominant mechanism
responsible for the different   brightness profiles.  }}

\subsection{M4 and NGC6397}

We have just summarised the main reason why our models of these two clusters
have distinctively different surface brightness profiles.  But we
pointed out in the previous section that this is not the same as
explaining why these two clusters have different profiles, partly because
there are so many aspects of their structure and evolution which we
have not considered in our modelling.  Here we list a few of these, and add some
remarks also on the role of the binary fraction and metallicity.

\subsubsection{Different tidal effects}

  It was pointed out by \citet{SC1973} that mild shocking of the halo
of a cluster by time-dependent tidal fields accelerates the collapse
of the core.  The mechanism for this is assumed to be the fact that
the shocking depletes the halo, leaving a greater volume of unoccupied
phase space into which high-energy core particles can  be ejected.
Severe shocking averts core collapse, but these authors concluded that
it was unlikely that any Galactic globular cluster surviving to the
present day was in this regime.

\citet{Di1999} find that the peri- and apogalactic distances for the
two clusters are about 0.6 and 5.9 kpc for M4, and 3.1 and 6.3 kpc for
NGC 6397.  These values agree quite well with other determinations
(\citet{AS1993} for M4; \citet{Mi2006}, \citet{Ka2007} for NGC 6397).  Thus M4
suffers the stronger and more frequent bulge shocks, and would be
expected to exhibit faster core collapse.  We are unable to quantify
this statement, however, as time-dependent tides cannot be included in
the modelling at present.  Nevertheless we note that the initial
concentration of our models of these clusters is very high, and it is
possible to argue that much of the evolution of their core is
relatively unaffected by the tide.  { At any rate, the faster
  evolution of M4 would seem to be unable to explain why it is the
  cluster which has the cored profile.}

\subsubsection{Different binary fractions}\label{sec:bf}

As already mentioned, different sources give rather discrepant
estimates of the binary fractions in these clusters, even when based
on the same technique, i.e. photometric offset binaries (Table
\ref{tab:fb}).  Nevertheless, the consensus is that the binary
fraction in NGC 6397 is not less than in M4.  Now it would be
expected that an increased binary fraction leads to a larger core
\citep[for example]{THH2007},
whereas it is clear that NGC 6397 has the smaller core.

\subsubsection{Different metallicity}

NGC6397 { has a much lower metallicity} than M4 and so, even with identical
initial conditions, the two clusters would evolve rather differently.
{ It is also known that the value of $M/L_V$ increases with
increasing metallicity in old populations \citep{Ma1999}.  It is not
at all clear, however, by what mechanism these effects should
primarily affect the {\sl central} surface brightness profile, and in
particular the core radius.  Core collapse is actually delayed in
low-metallicity cluster models, at least those with an initial mass of
a few times $10^4 M_\odot$ \citep{Hu2004}.  Among the reasons for this
is the fact that evolutionary time scales are shorter for lower
metallicity, and this results in a smaller mean mass and a greater
expansion (because of mass loss from stellar evolution) of systems at
lower metallicity.  In any event, it is again therefore surprising
that it is NGC 6397 which has the small core.  Equally inexplicable
is the}
marginal empirical evidence \citep{CD1989} that { so-called} ``core-collapse'' clusters
have lower metallicity than King-model clusters.

\subsubsection{Primordial mass segregation}

A topic of some recent dynamical interest is primordial mass
segregation, which is entirely absent from our modelling.  This is
related to the issue of surface brightness profiles through the
following chain of argument.  \citet{dempp2007} noticed that very
concentrated clusters tend to have steeper low-mass mass functions at
the present day than unconcentrated clusters, whereas if concentrated
clusters have undergone core collapse, we would expect them to be the
dynamically older group, to have lost relatively more low-mass stars,
and therefore to have the flatter present-day mass function.
\citet{Ba2008} argued, on the basis of $N$-body computations, that
this could be accounted for if one adopts a fixed initial mass
function, but with primordial mass segregation.  We claim that our
models of M4 and NGC 6397 are able, approximately, to account for both
the concentration and the present-day mass function, without
primordial mass segregation, but at the cost of allowing the initial
mass function to vary from the form favoured by \citet{Ba2008}.
Perhaps these assumptions (no initial mass segregation, and a fixed
initial mass function) are to some extent interchangeable, as far as
the effects on mass loss and the surface brightness profile are
concerned; { in an initially mass-segregated cluster, the early
rapid loss of stars tends to remove stars of low mass, thereby
flattening the mass function.}

\subsubsection{Other factors}

The only aspect of the initial mass function with which we have
experimented is the slope of the lower mass function.  Nevertheless it
is possible that other differences (e.g.the slope of the upper mass
function, the maximum mass, etc.) may play a role.

 { An aspect of
  stellar evolution which we have ignored is the possibility of natal
  kicks to black holes and/or white dwarfs \citep{Da2008b,MS2009,Fr2009},
  though we have shown (Sec.\ref{sec:nskicks}) that the effects of
  kicks in neutron stars on the observational data considered in this
  paper are minor.

  There are 
structural aspects which we have ignored, such as the role of cluster
rotation, which \citet{Ge1995} reported for NGC 6397.  Depending on
the mass function, this may or may not accelerate the process
of core collapse \citep{Ki2004}.  Equally, we have ignored some tidal
effects such as that due to encounters with spiral arms or giant
molecular clouds \citep{Gl2006,Gi2007}; these are thought to play a
minor role because of the high-inclination orbits on which most
globular clusters move.
}

\subsection{The initial conditions}

While our initial model for M4 was already surprisingly compact, our
initial conditions for NGC6397 are more compact still
(Tab.\ref{tab:ic-values}).   { As discussed in Sec.\ref{sec:init-conc}, }this seems to play a minor role in the
determination of the surface brightness profile, but it  raises once again
the question of how plausible
these { initial conditions} are.  On the other hand recent observations of nearby young massive star
clusters yield roughly comparable radii.  For example, \citet{BG2008} describe
clusters in the galaxy M51 with core radii of order 0.4pc, ages around
5Myr and masses of order $10^5M_\odot$.  While our initial half-mass
radius is the same, this implies that our initial core radius (in
the sense of the radius at which the projected mass density falls to
half its central value) is only 0.2pc.  Furthermore these authors
{ mention} that their interpretation of their observations assumes that
mass follows light.

\citet{BG2008} also point out that the early evolution of the core is
affected by three physical processes.  Of these, only two are included
in our models (mass loss from stellar evolution, and the settling of
black holes by mass segregation); but we omit the effect of residual
gas expansion.  \citet{Kr2008} has reviewed the early effects of this
process on such factors as the core radius.  What we have to assume is
that our models compensate this omission by an alteration in the
initial conditions, and that, at a time when the effects of this
process are over, our models have a structure comparable with that of
a cluster which {\sl is} subject to residual gas expulsion but started with
somewhat different initial conditions.  In much the same way, we note
that our models, by their compactness, experience an early phase of
extremely rapid mass segregation, and this may compensate the absence
of mass segregation in the initial conditions.  { Indeed the time
  scale for mass segregation in our initial model may be estimated to
  be about 0.5Myr\footnote{This was calculated by multiplying the half-mass
  relaxation time \citep{Sp1987} by the ratio of the mean mass divided
  by the maximum mass (50$M_\odot$).}.

This estimate for the time of mass segregation is considerably shorter
than in models of dense young stellar clusters in which runaway
coallescence has been found to occur \citep{PZ2004}.  But these
authors point out that the initial concentration, $c$, i.e. the
logarithm of the ratio of tidal to
core radius, is also important: 
 their results suggest that we require $c \gtorder 2$ for runaway
coallescence when the mass segregation time scale is 3Myr.  While our
initial Plummer model is more concentrated, it is not the ratio of
core to tidal radius which matters; in terms of the ratio of core to
{\sl half-mass} radius, our initial models are considerably less
concentrated.  Indeed, } runaway
coallescence is not a particularly striking feature of the results,
the largest stellar masses achieved being just over 150$M_\odot$,
{ i.e. three times the initial maximum mass}.  

\subsection{Fluctuations}

{ An interesting problem for the future is an exploration of the
  dynamical mechanism underlying the fluctuations we observe in the
  models.  It is tempting to think of these as analogous to the
  fluctuations caused by the motions of stars in some fixed potential.
  But it is also possible that they are more akin to incipient
  gravothermal oscillations.  \citet{Mu1990} discussed the occurrence
  of gravothermal oscillations in a multi-mass Fokker-Planck model
  with a mass-range not grossly dissimilar to our evolved models of M4
  and NGC6397, and found that gravothermal oscillations become
  apparent when the total mass exceeds about $8\times10^4M_\odot$.
  While this is close to the total mass of our evolved models (Table
  \ref{tab:ic-values}), we should recall that Fokker-Planck models by
  \citet{TI1991} show that even a core which is stable to gravothermal
  oscillations exhibits large fluctuations if the stochastic nature of
  binary heating is modelled.  We suggest that this is the explanation
  of the rather significant, irregular oscillations in the central
  escape speed (Fig.\ref{fig:escape-speed}), which are particularly
  noticeable after core collapse.}

Our focus on determining the initial conditions has actually been
somewhat undermined by our findings.  We have concluded that the
structure of a rich star cluster, after 12 Gyr of evolution, is a
product not just of the initial conditions but also of fluctuations.
{    The best one can do is to say that certain
initial conditions give rise to a statistical distribution of
outcomes.  In other words { (though the physics and timescales are
totally different)} the problem is not so different from
forecasting the weather.  If we forecast a day of sunshine and showers, we cannot
predict, for a given time,  whether sun will shine or rain will fall.
In the same way, while we may ``predict'' that a given cluster is a
post-collapse cluster, we cannot say whether or not it has a
resolvable core.}
Therefore even within the simplified set of initial conditions which
we have explored, there can be no {\sl unique} choice which leads to a
model with a structure and dynamics matching any given globular star
cluster at the present day.

Much attention has been paid in the literature to understanding how
the size of a post-collapse core depends on the binary fraction and
other factors, as if this were a deterministic question.  The results
of this paper imply that equal attention should be paid to the
fluctuations in the core radius, and how their statistics depend on
the mass function, the binary fraction, and so on.

\section*{Acknowledgements}

We are indebted to A. Milone for unpublished information on the binary
population, B. Hansen for information on the white dwarf luminosity
function (though we have not discussed this here), 
and to G. de Marchi for help on some photometric issues.
{ J. Hurley has patiently and quickly responded to our enquiries about
  his stellar evolution package.}  { We received extensive and very
  useful comments from S.F. Portegies Zwart on the previous version of
  the paper, and these have been most helpful.}
This research was supported in part by the Polish National Committee
for Scientific Research under grant 1 P03D 002 27, { and in part 
by the Polish Ministry of Science and Higher
Education through the grant 92/N--ASTROSIM/2008/0. MIG warmly thanks
DCH for his hospitality during a visit to Edinburgh which gave a boost to
the project.}  The work reported
at the end of Sec.\ref{sec:params} is based on extensive data
processing by Grzegorz Wiktorowicz, { who was supported by the Student
Summer Programme at CAMK.}

\bsp

\label{lastpage}


\begin{thebibliography}{99}

\bibitem[\protect\citeauthoryear{Allen 
\& Santillan}{1993}]{AS1993} Allen C., Santillan A., 1993, RMxAA, 25, 39 
\bibitem[\protect\citeauthoryear{Auriere, Lauzeral, 
\& Ortolani}{1990}]{Au1990} Auriere M., Lauzeral C., Ortolani S., 1990, Nature, 344, 638 

\bibitem[\protect\citeauthoryear{Baraffe et 
al.}{1997}]{Ba1997} Baraffe I., Chabrier G., Allard F., Hauschildt P.~H., 1997, A\&A, 327, 1054 


\bibitem[\protect\citeauthoryear{Bastian et 
al.}{2008}]{BG2008} Bastian N., Gieles M., Goodwin S.~P., 
Trancho G., Smith L.~J., Konstantopoulos I., Efremov Y., 2008, MNRAS,
389, 
223 

\bibitem[\protect\citeauthoryear{Baumgardt, De Marchi, 
\& Kroupa}{2008}]{Ba2008} Baumgardt H., De Marchi G., Kroupa P., 2008, ApJ, 685, 247 



\bibitem[\protect\citeauthoryear{Chernoff 
\& Djorgovski}{1989}]{CD1989} Chernoff D.~F., Djorgovski S., 1989, ApJ, 339, 904 



\bibitem[\protect\citeauthoryear{Chernoff 
\& Weinberg}{1990}]{CW1990} Chernoff D.~F., Weinberg M.~D., 1990, ApJ, 351, 121 




\bibitem[\protect\citeauthoryear{Davis et al.}{2008a}]{Da2008a} 
Davis D.~S., Richer H.~B., Anderson J., Brewer J., Hurley J., Kalirai 
J.~S., Rich R.~M., Stetson P.~B., 2008a, AJ, 135, 2155 

\bibitem[\protect\citeauthoryear{Davis et al.}{2008b}]{Da2008b} 
{ Davis D.~S., Richer H.~B., King I.~R., Anderson J., Coffey J., Fahlman 
G.~G., Hurley J., Kalirai J.~S., 2008b, MNRAS, 383, L20 }




\bibitem[\protect\citeauthoryear{De Marchi, Paresce, \& 
Pulone}{2007}]{dempp2007} De Marchi G., Paresce F., Pulone L., 
2007, ApJ, 656, L65 


\bibitem[Dinescu et al.(1999)]{Di1999} Dinescu, D.~I., Girard, 
T.~M., \& van Altena, W.~F.\ 1999, \aj, 117, 1792





\bibitem[\protect\citeauthoryear{Drukier}{1993}]{Dr1993} 
Drukier G.~A., 1993, MNRAS, 265, 773 

\bibitem[\protect\citeauthoryear{Drukier}{1995}]{Dr1995} 
Drukier G.~A., 1995, ApJS, 100, 347 

\bibitem[\protect\citeauthoryear{Fregeau et al}{2009}]{Fr2009}
{ Fregeau J.M., Richer H.B., Rasio F.A., Hurley J.R., 2009,  arXiv:0902.1166v1 }


\bibitem[\protect\citeauthoryear{Gebhardt et 
al.}{1995}]{Ge1995} { Gebhardt K., Pryor C., Williams T.~B., 
Hesser J.~E., 1995, AJ, 110, 1699 }

\bibitem[\protect\citeauthoryear{Gieles et al.}{2006}]{Gl2006} 
{ Gieles M., Portegies Zwart S.~F., Baumgardt H., Athanassoula E., Lamers 
H.~J.~G.~L.~M., Sipior M., Leenaarts J., 2006, MNRAS, 371, 793 }

\bibitem[\protect\citeauthoryear{Gieles, Athanassoula, 
\& Portegies Zwart}{2007}]{Gi2007} { Gieles M., Athanassoula E., Portegies Zwart S.~F., 2007, MNRAS, 376, 809 
}






\bibitem[Giersz(1998)]{Gi1998} Giersz, M.\ 1998, \mnras, 298, 
1239 

\bibitem[Giersz(2001)]{Gi2001} Giersz, M.\ 2001, \mnras, 324, 
218 

\bibitem[Giersz(2006)]{Gi2006} Giersz, M.\ 2006, \mnras, 371, 
484 




\bibitem[\protect\citeauthoryear{Giersz 
\& Heggie}{2003}]{GH2003} Giersz M., Heggie D.~C., 2003, MNRAS, 339, 486 



\bibitem[\protect\citeauthoryear{Giersz, Heggie, 
\& Hurley}{2008}]{GHH2008} Giersz M., Heggie D.~C., Hurley
  J.~R., 2008, MNRAS, 388, 429 

\bibitem[\protect\citeauthoryear{Hansen et al.}{2004}]{Ha2004} 
{ Hansen B.~M.~S., et al., 2004, ApJS, 155, 551 }


\bibitem[\protect\citeauthoryear{Hansen et al.}{2007}]{Ha2007} 
{ Hansen B.~M.~S., et al., 2007, ApJ, 671, 380 }




\bibitem[Harris(1996)]{Ha1996} Harris, W.~E.\ 1996, \aj, 112, 
1487

\bibitem[\protect\citeauthoryear{Heggie 
\& Giersz}{2008}]{HG2008} Heggie D.~C., Giersz M., 2008,
  MNRAS, 389, 1858



\bibitem[\protect\citeauthoryear{H{\'e}non}{1971}]{He1971} H{\'e}non M.~H., 1971, Ap\&SS, 14, 151 




 \bibitem[\protect\citeauthoryear{Hurley}{2007}]{Hu2007} Hurley 
 J.~R., 2007, MNRAS, 379, 93 



\bibitem[\protect\citeauthoryear{Hurley et al.}{2004}]{Hu2004} 
Hurley J.~R., Tout C.~A., Aarseth S.~J., Pols O.~R., 2004, MNRAS, 355, 1207 






\bibitem[\protect\citeauthoryear{Hurley et al.}{2008}]{Hu2008} 
Hurley J.~R., et al., 2008, AJ, 135, 2129 



\bibitem[\protect\citeauthoryear{Kalirai et 
al.}{2007}]{Ka2007} Kalirai J.~S., et al., 2007, ApJ, 657, L93 

\bibitem[\protect\citeauthoryear{Kim, Lee, 
\& Spurzem}{2004}]{Ki2004} { Kim E., Lee H.~M., Spurzem R., 2004, MNRAS, 351, 220 
}


 \bibitem[Kroupa(1995)]{Kr1995} Kroupa, P.\ 1995, \mnras, 277, 
 1507 



\bibitem[\protect\citeauthoryear{Kroupa}{2008}]{Kr2008}
  Kroupa 
P., 2008,  arXiv:0803.1833 





\bibitem[\protect\citeauthoryear{Maraston}{1999}]{Ma1999} 
Maraston C., 1999, ASPC, 163, 28 



\bibitem[\protect\citeauthoryear{Meylan 
\& Mayor}{1991}]{MM1991} Meylan G., Mayor M., 1991, A\&A, 250, 113



\bibitem[\protect\citeauthoryear{Milone et 
al.}{2006}]{Mi2006} Milone A.~P., Villanova S., Bedin L.~R., Piotto G., Carraro G., Anderson J., King I.~R., Zaggia S., 2006, A\&A, 456, 517 

\bibitem[\protect\citeauthoryear{Milone et al.}{2008}]{Mi2008} 
Milone A.~P., Piotto G., Bedin L.~R., Sarajedini A., 2008, MmSAI, 79, 623 

\bibitem[\protect\citeauthoryear{Moody 
\& Sigurdsson}{2009}]{MS2009} {\sl Moody K., Sigurdsson S., 2009, ApJ, 690, 1370 }




\bibitem[\protect\citeauthoryear{Mould et al.}{1996}]{Mo1996} 
Mould J.~R., et al., 1996, PASP, 108, 682

\bibitem[\protect\citeauthoryear{Murphy, Cohn, 
\& Hut}{1990}]{Mu1990} Murphy B.~W., Cohn H.~N., Hut P., 1990, MNRAS, 245, 335 


\bibitem[\protect\citeauthoryear{Noyola 
\& Gebhardt}{2006}]{NG2006} Noyola E., Gebhardt K., 2006, AJ, 132, 447


\bibitem[\protect\citeauthoryear{Paresce, de Marchi, 
\& Romaniello}{1995}]{Pa1995} Paresce F., de Marchi G., Romaniello M., 1995, ApJ, 440, 216


\bibitem[\protect\citeauthoryear{Pfahl, Rappaport, 
\& Podsiadlowski}{2002}]{Pf2002} { Pfahl E., Rappaport S., Podsiadlowski P., 2002, ApJ, 573, 283}


\bibitem[\protect\citeauthoryear{Portegies Zwart et 
al.}{2004}]{PZ2004} { Portegies Zwart S.~F., Baumgardt H., Hut 
P., Makino J., McMillan S.~L.~W., 2004, Nature, 428, 724; astro-ph/arXiv:astro-ph/0402622v1
}









\bibitem[Richer et al.(2004)]{Ri2004} Richer H.~B., et al.,\ 
2004, \aj, 127, 2771 

\bibitem[\protect\citeauthoryear{Richer et al.}{2008}]{Ri2008} 
Richer H.~B., et al., 2008, AJ, 135, 2141 



\bibitem[\protect\citeauthoryear{Sommariva et 
al.}{2008}]{So2008} Sommariva V., Piotto G., Rejkuba M., Bedin 
L.~R., Heggie D.~C., Mathieu R.~D., Villanova S., 2009, A\&A, 493, 947


\bibitem[\protect\citeauthoryear{Spitzer}{1987}]{Sp1987} 
{ Spitzer L., 1987, {\sl Dynamical Evolution of Globular Clusters},  Princeton: University Press}




\bibitem[\protect\citeauthoryear{Spitzer 
\& Chevalier}{1973}]{SC1973} Spitzer L.~J., Chevalier R.~A., 1973, ApJ, 183, 565 




\bibitem[\protect\citeauthoryear{Stodo\l kiewicz}{1982}]{St1982} 
Stodo\l kiewicz J.~S., 1982, AcA, 32, 63 

\bibitem[\protect\citeauthoryear{Stodo\l kiewicz}{1986}]{St1986} 
Stodo\l kiewicz J.~S., 1986, AcA, 36, 19 










\bibitem[\protect\citeauthoryear{Sugimoto 
\& Bettwieser}{1983}]{SB1983} Sugimoto D., Bettwieser E., 1983, MNRAS, 204, 19P 

\bibitem[\protect\citeauthoryear{Takahashi 
\& Inagaki}{1991}]{TI1991} Takahashi K., Inagaki S., 1991, PASJ, 43, 589 




 \bibitem[Trager et al.(1993)]{Tr1993} Trager, S.~C., 
 Djorgovski, S., \& King, I.~R.\ 1993, in Djorgovski,
   S.G., Meylan G., eds, Structure and Dynamics of Globular 
 Clusters, ASPCS 50, 347

\bibitem[Trager et al.(1995)]{Tr1995} Trager, S.~C., King, 
I.~R., \& Djorgovski, S.\ 1995, \aj, 109, 218

\bibitem[\protect\citeauthoryear{Trenti, Heggie, 
\& Hut}{2007}]{THH2007} Trenti M., Heggie D.~C., Hut P., 2007, MNRAS, 374, 344 



\end{thebibliography}
\end{document}